\documentclass[pra, showpacs, twocolumn, floatfix,superscriptaddress]{revtex4}
\usepackage{graphicx}
\usepackage{epstopdf}
\usepackage{amsmath, amsfonts, amssymb, bm}

\def\mpik{Max-Planck-Institut f\"ur Kernphysik, Saupfercheckweg 1, D-69117
Heidelberg, Germany}

\def\ifa{Institute of Applied Physics, Academy of Sciences of Moldova,
Academiei str. 5, MD-2028 Chi\c{s}in\u{a}u, Moldova}
\begin{document}
\title{Population inversion in two-level systems possessing permanent dipoles}
\author{Mihai  \surname{Macovei}}
\email{macovei@phys.asm.md}
\affiliation{\mpik}
\affiliation{\ifa}

\author{Mayank \surname{Mishra}}
\thanks{Present address: IISER Mohali, Knowledge city, Sector 81, SAS Nagar, Manauli, Punjab, India, PO 140306.}

\author{Christoph H. \surname{Keitel}}
\affiliation{\mpik}

\date{\today}
\begin{abstract}
Bare-state population inversion is demonstrated in a two-level system with all dipole matrix elements nonzero. A laser field is resonantly driving 
the sample whereas a second weaker and lower frequency coherent field additionally pumps it near resonance with the dynamically-Stark-splitted states.  
Due to existence of differing permanent dipole moments in the excited and ground bare states, quantum coherences among the involved 
dressed-states are induced leading to inversion in the steady-state. Furthermore, large refractive indices are feasible as well as the determination of 
the diagonal matrix elements via the absorption or emission spectra. The results apply to available biomolecular, spin or asymmetric quantum dot systems. 
\end{abstract}
\maketitle
\section{Introduction}
Population inversion in a few-energy-level quantum system is strongly linked with its optical response and has resulted in traditional masing or lasing effects 
being successfully demonstrated \cite{townes,prohorov,basov}. Subsequently, enormous economical and technological progress was triggered due to 
quantum electronics. However at present, atomic steady-state population inversion is achievable essentially only in pumped two-level atomic systems involving 
efforts via extra transitions with additional photon sources or environmental vacuum modifications \cite{inv_at}. Additionally, two-level quantum dot systems 
may also exhibit population inversion due to extra phonon induced decay rates \cite{inv_qd}. Meanwhile, lasing without population inversion operates as well 
with the help of induced quantum coherences \cite{lwi}. Moreover, free-electron laser sources involving inversions of momentum states exist at higher 
frequencies leading to original effects \cite{joerg}. 

Over the past decade a range of experiments were performed demonstrating quantum effects in biological samples \cite{q_bio}. Particularly, energy transfer 
through quantum coherence in photosynthetic systems \cite{q_coh} was observed in \cite{q_tr1,q_tr2}, while long-living quantum coherences survive in 
biological complexes even under normal conditions at room temperature \cite{lq_coh}. Quantum coherence and entanglement in the processes of 
magneto-reception of the surrounding magnetic field \cite{exp_rec,plenio} were examined, too~\cite{m_rec}. The single molecule fluorescence spectroscopy 
and the emitted photon quantum statistics are further excellent tools for research in quantum biology \cite{sms,ph_st}. Furthermore, coherent control of an 
effective two-level system in a non-Markovian biomolecular environment was investigated~\cite{nj}. Additional quantum effects in biochemical systems are 
discussed, for instance, in \cite{kov}. An important issue raised in this context is to image tiny objects, such as biological cells or organic molecules and, 
therefore, highly refractive biological media are required \cite{koch}.

Inspired by these remarkable advances in quantum biochemistry, here, we put forward a novel scheme that enables the creation of population inversion 
in certain biological samples acting as two-level systems with all possible dipole matrix elements being nonzero. The effect occurs due to induced 
quantum coherences which arise from the difference of the permanent dipole moments in the excited and ground states, respectively. 
This may allow for lasing or amplifying as well as optical switching devises in biomolecular materials. Large index of refractions without 
absorption are also feasible which may lead to an enhanced optical imaging resolution of the biomolecular sample due to a reduced probe-field wavelength 
inside the medium. Both the absorption or emission spectra can be used to extract the values of permanent dipoles.

The article is organized as follows. In Sec. II we describe
the analytical approach and the system of interest, whereas in
Sec. III we analyze the obtained results. A summary is given
in Sec. IV.

\section{Theoretical framework}
\begin{figure}[t]
\includegraphics[width =8cm]{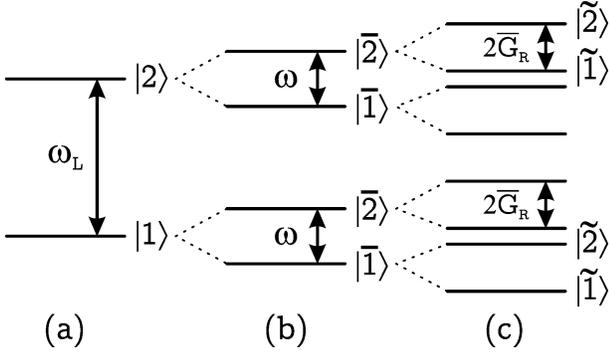}
\caption{\label{fig} 
Energy diagram of a two-level emitter with nonzero  values of all involved dipole matrix elements $d_{\alpha\beta}$, $\{\alpha, \beta =1,2 \}$. 
(a) A moderately intense laser field of frequency $\omega_{L}$ resonantly interacts with the molecular sample leading to 
dynamically Stark splitting of its energy levels. (b) A second coherent source of frequency $\omega$, close to generalized Rabi frequency due to 
first laser, is applied leading to transitions among the dressed-states. (c) The "double dressed-states" with $2\bar G_{R}$ being the corresponding 
Rabi splitting frequency.}
\end{figure}

We consider a two-level system possessing permanent dipoles and interacting with two external coherent laser fields. The first laser is near resonance with 
the transition frequency of the two-level sample while the second one is close to resonance with the dressed-frequency splitting due to the first laser, 
respectively (see Figure \ref{fig}). As a concrete system, we may consider gamma-globulin macromolecules \cite{kov,setl} with the transition frequency
$\omega_{21} \approx 4.8\times 10^{15}{\rm Hz}$, transition dipole moment $d \approx {\rm 1Debye}$ and the difference between the diagonal dipole moments in the 
upper and lower bare states given by $|d_{22}-d_{11}| \approx {\rm 100Debye}$. However, the analytical formalism 
applies equally to spin \cite{spin}, asymmetric semiconductor quantum dot \cite{q_dots} or other alternative \cite{gong} systems promising wider applications. 
The Hamiltonian describing such a model, in a frame rotating at the first laser frequency $\omega_{L}$, and in the dipole approximation is:
\begin{eqnarray}
H&=& \sum_{k}\hbar(\omega_{k}-\omega_{L})a^{\dagger}_{k}a_{k} + \hbar\Delta S_{z} + \hbar\Omega (S^{+}+S^{-}) \nonumber \\
&+&  \hbar G S_{z}\cos{(\omega t )} + i\sum_{k}(\vec g_{k}\cdot \vec d)(a^{\dagger}_{k}S^{-} - a_{k}S^{+}). \label{Ham}
\end{eqnarray}
In the Hamiltonian (\ref{Ham}) the first three components are, respectively, the free energies of the environmental electromagnetic vacuum modes and 
molecular subsystems together with the laser-molecule interaction Hamiltonian. There, $\Omega = d E_{1}/(2\hbar)$ is the corresponding Rabi frequency 
with $d \equiv d_{21}=d_{12}$ being the transition dipole moment while $E_{1}$ is the amplitude of the first laser field. 
The fourth term accounts for the second laser interacting at frequency $\omega$ and amplitude $E_{2}$ with the molecular system due to presence 
of permanent dipoles incorporated in $G$, i.e., $G=(d_{22}-d_{11})E_{2}/\hbar$. 
The last term describes the interaction of the molecular subsystem with the environmental vacuum modes of the electromagnetic field reservoir. Further,  $\vec g=\sqrt{2\pi\hbar\omega_{k}/V}\vec e_{\lambda}$ is the molecule-vacuum coupling strength with 
$\vec e_{\lambda}$ being the photon polarization vector and $\lambda \in \{1,2\}$ whereas $V$ is the quantization volume. $\Delta=\omega_{21}-\omega_{L}$ 
is the laser field detuning from the molecular transition frequency $\omega_{21}$. The molecule bare-state operators $S^{+}=|2\rangle \langle 1|$ and
$S^{-}=[S^{+}]^{\dagger}$ obey the commutation relations $[S^{+},S^{-}]=2S_{z}$ and $[S_{z},S^{\pm}]=\pm S^{\pm}$. 
Here, $S_{z}=(|2\rangle \langle 2|-|1\rangle \langle 1|)/2$ is the bare-state inversion operator. $|2\rangle$ and $|1\rangle$ are, 
respectively, the excited and ground state of the molecule while $a^{\dagger}_{k}$ and $a_{k}$ are the creation and the annihilation
operator of the $k_{th}$ electromagnetic field mode, and satisfy the standard bosonic commutation relations, namely, 
$[a_{k},a^{\dagger}_{k^{'}}]=\delta_{kk^{'}}$, and $[a_{k},a_{k^{'}}]=[a^{\dagger}_{k},a^{\dagger}_{k^{'}}]=0$ \cite{gsa_book,al_eb,puri,martin}. 
Notice that the Hamiltonian (\ref{Ham}) is incomplete. The following term 
\begin{eqnarray}
\tilde H = \hbar \tilde \Omega S^{+}e^{i\omega_{L}t}\cos{(\omega t)} + \hbar \tilde G S_{z}\cos{(\omega_{L}t)}/2 + H.c., \label{hamm}
\end{eqnarray}
will not be taken into account. Here, the first term describes the interaction of the second laser with the molecular system due to the transition dipole $d$, while 
the second one represents the interaction of the first laser with the molecule because of permanent dipoles and, hence, $\tilde \Omega=d E_{2}/\hbar$ 
whereas $\tilde G = (d_{22}-d_{11}) E_{1}/\hbar$.  Since we assume realistic conditions that $\tilde \Omega \ll \omega_{L} \pm \omega$ \cite{puri,martin} as 
well as $\{\tilde G, \omega\} \ll \omega_{L}$ the Hamiltonian $(\ref{hamm})$ can be considered as fast oscillating and, therefore, neglectable. 

In the following, we shall consider a regime where the generalized Rabi frequency $\bar \Omega = \sqrt{\Omega^{2}+ (\Delta/2)^{2}}$ is larger than the 
single-molecule spontaneous decay rate as well as the coupling due to permanent dipoles, i.e. $\bar \Omega \gg \gamma$ and $\bar \Omega > G$. 
In this case it is more convenient to describe our system in the semi-classical laser-molecule dressed-state picture due to the first applied laser: 
\begin{eqnarray}
|2\rangle = \cos{\theta}|\bar 2\rangle -\sin{\theta}|\bar 1\rangle ~ {\rm and}~
|1\rangle = \cos{\theta}|\bar1\rangle + \sin{\theta}|\bar 2\rangle, 
\label{drs1}
\end{eqnarray}
with $\tan{2\theta}=2\Omega/\Delta$. Here $|\bar 2\rangle$ and $|\bar 1\rangle$ are the corresponding upper and lower dressed states, respectively 
(see Figure~\ref{fig}b). 
Applying the dressed-state transformation to the Hamiltonian (\ref{Ham}) one arrives at the following Hamiltonian represented in a frame rotating also 
at the second laser field frequency, i.e. $\omega$,
\begin{eqnarray}
H&=& \sum_{k}\hbar(\omega_{k}-\omega_{L})a^{\dagger}_{k}a_{k} + \hbar\bar\Delta R_{z} - \hbar\bar G (R^{+} + R^{-}) \nonumber \\
&+& i\sum_{k}(\vec g_{k}\cdot \vec d)\{ a^{\dagger}_{k}(\sin{2\theta}R_{z}/2 + \cos^{2}{\theta}R^{-}e^{-i\omega t} \nonumber \\
&-& \sin^{2}{\theta}R^{+}e^{i\omega t}) - H.c.\}, \label{Hamd}
\end{eqnarray}
where $\bar \Delta = \bar \Omega - \omega(1-\bar G^{2}/\omega^{2})/2 \equiv \bar \Omega - \omega/2$ and $\bar G=(G/4)\sin{2\theta}$, and 
we have performed the rotating wave approximation with respect 
to $\omega$, i.e., we have assumed that $\omega \gg \bar G$. Eliminating the vacuum modes of the electromagnetic field reservoir in the usual way by 
adopting the Born-Markov approximations \cite{gsa_book,al_eb,puri,martin} one arrives then at the following dressed-state master equation:
\begin{eqnarray}
& \dot \rho(t) + i[\bar\Delta R_{z} - \bar G(R^{+} + R^{-}),\rho] = -\frac{\gamma_{0}}{4}\sin^{2}{2\theta}[ R_{z},R_{z}\rho] \nonumber \\
& -\gamma_{+}\cos^{4}{\theta}[R^{+},R^{-}\rho] - \gamma_{-}\sin^{4}{\theta}[R^{-},R^{+}\rho] +  H.c. \label{meq}
\end{eqnarray}
Here $\gamma_{0,\pm}$ are the single-qubit spontaneous decay rates corresponding to dressed-state frequencies $\omega_{L}$ and 
$\omega_{L} \pm 2\bar \Omega$, respectively.
The new quasi-spin operators, i.e., $R^{+}=|\bar 2\rangle\langle \bar 1|$, $R^{-}=[R^{+}]^{\dagger}$ 
and $R_{z}=|\bar 2\rangle\langle \bar 2| - |\bar 1\rangle\langle \bar 1|$ are operating in the dressed-state picture. They obey the following commutation 
relations: $[R^{+},R^{-}]=R_{z}$ and $[R_{z},R^\pm]=\pm 2R^\pm$. Notice that in the above master equation, we have neglected the rapidly oscillating 
terms in the spontaneous emission part - an approximation valid when $\{\bar \Omega, \omega \} \gg \{\gamma_{0},\gamma_{\pm}\}$.

In the following section, we will discuss our results, i.e., the possibility to create bare-state population inversion as well as high refractive media.
\section{Results and discussion}
One can observe from Eq.~(\ref{meq}) that permanent dipoles lead to the appearance of a pumping term among the dressed states. This term contributes 
to a completely unexpected behavior of the two-level system. In particular, as we shall see, it can induce pumping the system into an inverted state. The 
system of equations for the dressed-state inversion and dressed-state polarization operators can be obtained from the master equation (\ref{meq}), namely,
\begin{eqnarray}
\langle \dot R_{z}\rangle &=& -2i\bar G(\langle R^{-}\rangle - \langle R^{+}\rangle) - 2\Gamma_{+}\langle R_{z}\rangle + 2\Gamma_{-}, \nonumber \\
\langle \dot R^{+}\rangle &=& (2i\bar\Delta - \Gamma)\langle R^{+}\rangle + i\bar G \langle R_{z}\rangle, 
\label{seq}
\end{eqnarray}
with $\langle R^{-}\rangle=[\langle R^{+}\rangle]^{\dagger}$. Here, $\Gamma_{\pm}$=$\gamma(\sin^{4}{\theta} \pm \cos^{4}{\theta})$, 
$\Gamma= \Gamma_{+}+\gamma\sin^{2}{2\theta}$ and we have 
considered that $\gamma_{0}=\gamma_{\pm} \equiv \gamma$ which is the case for a free-space setup.
The mean-value of the bare-state inversion operator $\langle S_{z}\rangle$ can be represented via dressed-state operators as follows:
\begin{eqnarray}
\langle S_{z}\rangle = \cos{2\theta}\langle R_{z}\rangle/2 - \sin{2\theta}(\langle R^{+}\rangle + \langle R^{-}\rangle)/2. 
\label{bin}
\end{eqnarray}
From the system of equations (\ref{seq}) one immediately obtains the steady-state relations:
\begin{eqnarray}
\langle R^{+}\rangle = i\bar G\langle R_{z}\rangle/(\Gamma - 2i\bar \Delta) ~~ {\rm and}~~
\langle R^{-}\rangle = [\langle R^{+}\rangle]^{\dagger}. \label{pm}
\end{eqnarray}
Inserting (\ref{pm}) in Eq.~(\ref{bin}) one arrives at:
\begin{eqnarray}
\langle S_{z}\rangle = \bigl(\cos{2\theta} + 4\bar\Delta\bar G\sin{2\theta}/[\Gamma^{2} + (2\bar\Delta)^{2}]\bigr)\langle R_{z}\rangle/2, \label{binn}
\end{eqnarray}
where, again, from Eqs.~(\ref{seq}) one has that:
\begin{eqnarray}
\langle R_{z}\rangle = 2\Gamma_{-}/[2\Gamma_{+}+(2\bar G)^{2}\Gamma/(\Gamma^{2}+(2\bar \Delta)^{2})]. \label{ss}
\end{eqnarray}
\begin{figure}[t]
\includegraphics[width =8.7cm,height=3.5cm]{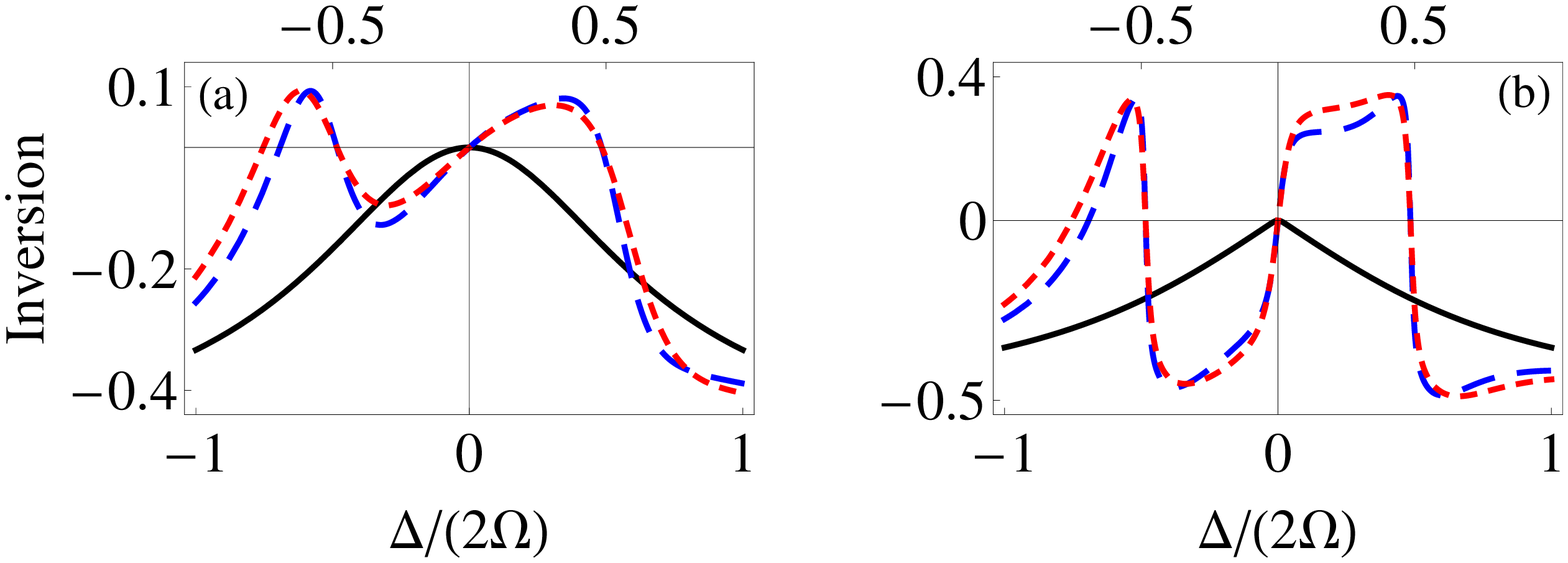}
\caption{\label{fig1} (color online) 
The steady-state dependence of the bare-state  inversion operator $\langle S_{z}\rangle/N$
versus the scaled parameter $\Delta/(2\Omega)$. The solid line is plotted for $G/\gamma=0$, the long-dashed one is for $G/\gamma=16$ 
while the short-dashed curve corresponds to $G/\gamma=24$. Other parameters are: $\Omega/(N\gamma) =45$ and $\omega/(N\gamma)=100$ 
with $\gamma \approx 2.6{\rm MHz}$ as feasible for gamma-globulin. (a) $N=1$ whereas (b) $N=50$ while molecules are dense enough to allow for collectivity.}
\end{figure}
An interesting result here is the non-zero value for the dressed-state coherences $\langle R^{\pm}\rangle$, see Eq.~(\ref{pm}). In the absence of 
permanent dipoles, i.e. $G=0$, these quantities are zero in the moderately intense pumping regime considered here. Therefore, new physics is expected 
due to existence of permanent dipoles in bichromatic pumping fields. In particular, Figure~{\ref{fig1}}(a) shows the mean-value of the single-molecule 
bare-state inversion operator for particular parameters of interest. Steady-state inversion in the bare states is achieved when $\langle S_{z}\rangle >0$ 
and it occurs in the  presence of permanent dipoles. On the other hand, Figure \ref{fig2}(a) depicts the real part of the mean-value of the dressed-state 
coherence operator $\langle R^{+}\rangle$ in steady-state. The minima observed in these behaviors correspond to an inverted molecular bare-state 
system (compare Fig.~\ref{fig1}a and Fig.~\ref{fig2}a). Thus, inversion occurs due to the real part of the dressed-state coherences which can be 
nonzero in our system.
\begin{figure}[b]
\includegraphics[width =8.7cm,height=3.5cm]{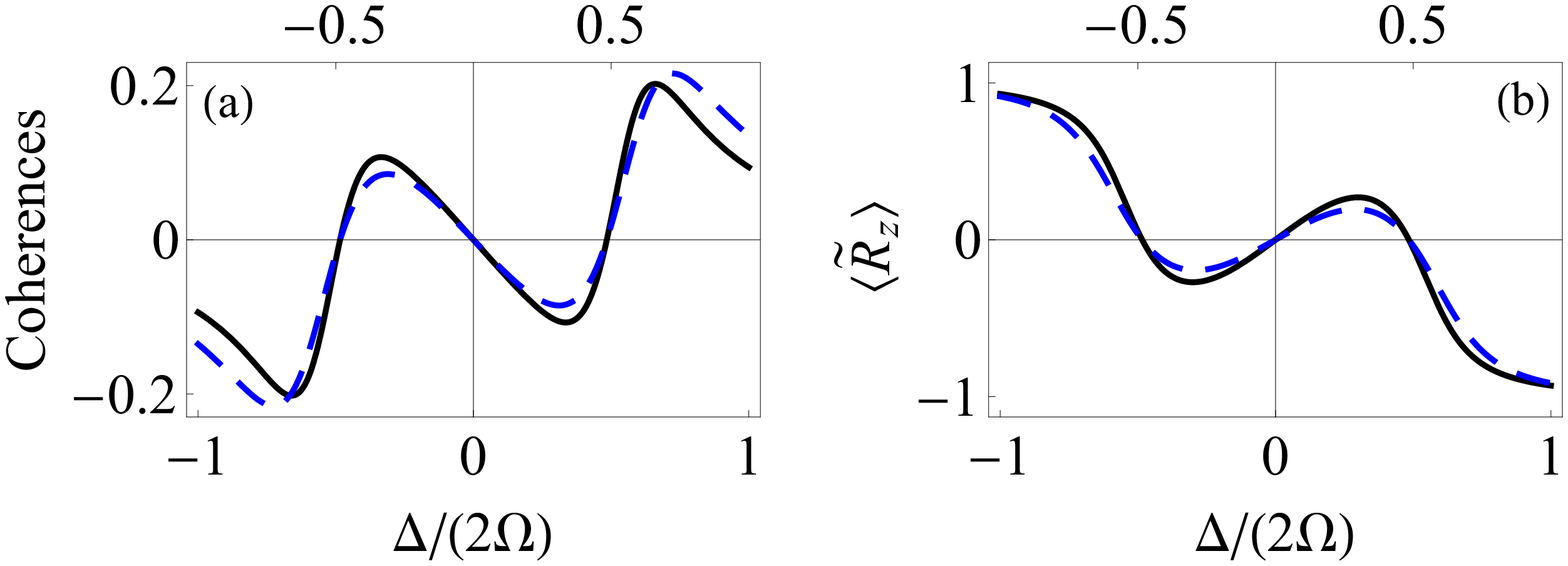}
\caption{\label{fig2} (color online) 
The single-molecule steady-state behaviors of (a) the real part of the dressed-state coherences $\langle R^{+}\rangle$  and (b) the double dressed-state 
inversion operator $\langle \tilde R_{z}\rangle$ versus the scaled detuning $\Delta/(2\Omega)$. The solid line is plotted for $G/\gamma=16$ 
while the dashed curve corresponds to $G/\gamma=24$. Here $\Omega/\gamma =45$ and $\omega/\gamma=100$.}
\end{figure}

One can apply the double dressed-state formalism (see Figure~\ref{fig}) in order to obtain further information on our system, namely,
\begin{eqnarray}
|\bar 2\rangle = \cos{\phi}|\tilde 2\rangle + \sin{\phi}|\tilde 1\rangle, ~
|\bar 1\rangle = \cos{\phi}|\tilde1\rangle - \sin{\phi}|\tilde 2\rangle. \label{drs2}
\end{eqnarray}
This approach is particularly useful to account for vacuum induced collective effects among the emitters and their corresponding influences on molecular 
dynamics. Introducing (\ref{drs2}) in the Hamiltonian (\ref{Hamd}) and, again, eliminating the degrees of freedom related with the environmental vacuum 
modes in the Born-Markov approximations one arrives at the double dressed master equation:
\begin{eqnarray}
\dot \rho(t) &+& i\bar G_{R}[\tilde R_{z},\rho] = -\bar \Gamma_{0}[\tilde R_{z}, \tilde R_{z}\rho] - \bar \Gamma_{+}[\tilde R^{+}, \tilde R^{-}\rho] 
\nonumber \\
&-& \bar \Gamma_{-}[\tilde R^{-}, \tilde R^{+}\rho] +  H.c. \label{meqd}
\end{eqnarray}
Here, $\bar \Gamma_{0}=\gamma(\omega_{L})\sin^{2}{2\theta}\cos^{2}{2\phi}/4 + \sin^{2}{2\phi}\{\gamma(\omega_{L}+\omega)\cos^{4}{\theta}
+\gamma(\omega_{L}-\omega)\sin^{4}{\theta}\}/4$, $\bar \Gamma_{+}=\gamma(\omega_{L}+2\bar G_{R})\sin^{2}{2\phi}\sin^{2}{2\theta}/4 + 
\gamma(\omega_{L}+\omega+2\bar G_{R})\cos^{4}{\phi}\cos^{4}{\theta} + \gamma(\omega_{L} - \omega + 2\bar G_{R})\sin^{4}{\theta}\sin^{4}{\phi}$ 
and
$\bar \Gamma_{-}=\gamma(\omega_{L}-2\bar G_{R})\sin^{2}{2\phi}\sin^{2}{2\theta}/4 + 
\gamma(\omega_{L}+\omega-2\bar G_{R})\cos^{4}{\theta}\sin^{4}{\phi} + \gamma(\omega_{L} - \omega - 2\bar G_{R})\sin^{4}{\theta}\cos^{4}{\phi}$ 
with $\cot{2\phi}=\bar \Delta/\bar G$, and $\bar G_{R}=\sqrt{\bar \Delta^{2}+\bar G^{2}}$.
The new operators, i.e., $\tilde R^{+}=|\tilde 2\rangle\langle \tilde 1|$, $\tilde R^{-}=[\tilde R^{+}]^{\dagger}$ 
and $\tilde R_{z}=|\tilde 2\rangle\langle \tilde 2| - |\tilde 1\rangle\langle \tilde 1|$ are operating in the double dressed-state picture obeying the following 
commutation relations: $[\tilde R^{+}, \tilde R^{-}] = \tilde R_{z}$ and $[\tilde R_{z}, \tilde R^\pm]=\pm 2\tilde R^\pm$. The master equation (\ref{meqd})
contains only slowly varying terms in the spontaneous emission damping, that is, we have assumed that $\bar G_{R} \gg \gamma(\tilde \omega)$, with
$\gamma(\tilde \omega)=2d^{2}\tilde \omega^{3}/(3\hbar c^{3})$ being the single-molecule spontaneous decay rate corresponding to the double 
dressed-state frequency $\tilde \omega$ (see Figure~\ref{fig}).

The steady-state solution of Eq.~(\ref{meqd}) can be chosen in the form:
\begin{eqnarray}
\rho = Z^{-1}\exp{[-\eta\tilde R_{z}]},
\label{dss}
\end{eqnarray}
where the normalization $Z$ is determined from the relation ${\rm Tr\{\rho\}=1}$. Inserting (\ref{dss}) into (\ref{meqd}) one obtains 
$\eta=\ln(\bar \Gamma_{+}/\bar \Gamma_{-})/2$. Using the relations (\ref{bin}) and (\ref{drs2}) we arrive at the following expression 
for the mean value of the bare-state inversion operator $\langle S_{z}\rangle$ represented via the double dressed-state inversion operator 
$\langle \tilde R_{z}\rangle$, respectively,
\begin{eqnarray}
\langle S_{z}\rangle = \cos{[2(\theta - \phi)]}\langle \tilde R_{z}\rangle/2.
\label{szz}
\end{eqnarray}
The steady-state expression for the double dressed-state inversion operator can be obtained with the help of
Eq.~(\ref{dss}) and the coherent molecular state $|s\rangle \equiv |N-s,s\rangle$ which denotes a symmetrized $N-$molecule state in which $N-s$ molecules 
are in the lower double dressed-state $|\tilde 1\rangle$ and $s$ molecules are excited to the upper double dressed-state $|\tilde 2\rangle$, respectively \cite{puri}. 
Thus, 
\begin{figure}[b]
\includegraphics[width =8.82cm,height=3.5cm]{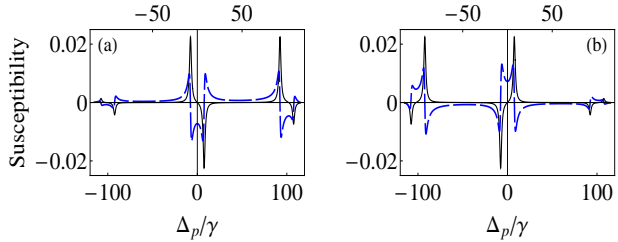}
\caption{\label{fig3} (color online) 
The steady-state dependence of the linear susceptibility $\chi(\nu_{p})$ [in units of $\bar N d^{2}/(\hbar \gamma)$] versus scaled detuning 
$\Delta_{p}/\gamma$. The solid black curve corresponds to the imaginary part (absorption spectrum) while the long-dashed blue line to the real 
part of the susceptibility, respectively. (a) $\Delta/(2\Omega)=0.43$ whereas (b) $\Delta/(2\Omega)=-0.43$. Other parameters are: 
$\Omega/\gamma=45$, $\omega/\gamma=100$, $G/\gamma=16$.}
\end{figure}
\begin{eqnarray}
\langle \tilde R_{z}\rangle = - N + \frac{(x^{1+N}-1)-(1+N)(x-1)}{2^{-1}(x-1)(x^{1+N}-1)},
\label{tdrs}
\end{eqnarray}
where $x=\bar \Gamma_{+}/\bar \Gamma_{-}$. In this case the molecular system has to occupy a volume with linear dimensions of the order of the 
smallest relevant emission wavelength or less. However, this restriction may be relaxed for certain geometries \cite{gsa_book,al_eb,puri,martin}. 
The molecular operators entering in Eqs.~(\ref{meqd}-\ref{tdrs}) are then collective ones, i.e. $\tilde R^{\pm}=\sum^{N}_{j=1}\tilde R^{\pm}_{j}$ 
as well as $\tilde R_{z}=\sum^{N}_{j=1}\tilde R_{zj}$. Figure {\ref{fig1}}(b) shows the bare-state inversion operator $\langle S_{z}\rangle/N$, based 
on the expressions (\ref{szz}) and (\ref{tdrs}), for a collection of $N=50$ molecules in a volume $(2\pi c/\omega_{21})^{3}$ or a molecular density 
$\bar N \approx 7.8\times 10^{14}{\rm cm^{-3}}$. Almost full inversion in the steady-state is achieved when we consider that the double-dressed 
decay rates $\gamma(\tilde \omega)$ are equal and denoted by $\gamma$. Furthermore, abrupt population behaviors are observed as well which may 
be used for engineering optical switching devices with switching times of the order of $(N\gamma)^{-1}$. Note that the mean values of non-diagonal terms 
resulting from Eq.~(\ref{meqd}) are zero in the steady-state.

We proceed by calculating the refractive properties of a very weak field probing the strongly driven molecular sample around the first laser's frequency
$\omega_{L}$. The linear susceptibility $\chi(\nu)$ of the probe field, at frequency $\nu$, can be represented in terms of the Fourier transform of the 
average value of the two-time commutator of the molecular operator as
\begin{eqnarray}
\chi(\nu) = \frac{i}{\hbar}\bar N d^{2}\int^{\infty}_{0}d\tau e^{i(\nu-\omega_{L})\tau}\langle[S^{-}(\tau),S^{+}]\rangle_{s}.
\label{sc}
\end{eqnarray}
Note that the subindex $s$ means steady-state.
Inserting the dressed-state transformations (\ref{drs1}) and (\ref{drs2}) in (\ref{sc}), in a frame rotating also at the second laser's frequency $\omega$, 
and using the master equation (\ref{meqd}) one arrives at the following expression for the susceptibility, namely,
\begin{eqnarray}
&{}&\chi(\nu)=\frac{i}{\hbar}\bar N d^{2}\langle \tilde R_{z}\rangle \bigl\{\frac{1}{4}\sin^{2}{2\theta}\sin^{2}{2\phi}\bigl(\chi_{1}(\Delta_{p},2\bar G_{R})
\nonumber \\
&-& \chi_{1}(\Delta_{p},-2\bar G_{R})\bigr) + \cos^{4}{\theta}\bigl(\sin^{4}{\phi}\chi_{2}(\Delta_{p},2\bar G_{R},-\omega) \nonumber \\
&-& \cos^{4}{\phi}\chi_{2}(\Delta_{p},-2\bar G_{R},-\omega)\bigr) + \sin^{4}{\theta}\bigl(\cos^{4}{\phi} \nonumber \\
&\times& \chi_{3}(\Delta_{p},2\bar G_{R},\omega) - \sin^{4}{\phi}\chi_{3}(\Delta_{p},-2\bar G_{R},\omega) \bigr) \bigr\}, \label{chidr}
\end{eqnarray}
where $\chi_{1}(\Delta_{p},x)=(\bar \Gamma_{s}+i(\Delta_{p}+x))/[\bar \Gamma^{2}_{s}+(\Delta_{p}+x)^{2}]$, 
$\chi_{2}(\Delta_{p},x,-y)=(\bar \Gamma_{s}+i(\Delta_{p}+x-y))/[\bar \Gamma^{2}_{s}+(\Delta_{p}+x-y)^{2}]$
and $\chi_{3}(\Delta_{p},x,y)=(\bar \Gamma_{s}+i(\Delta_{p}+x+y))/[\bar \Gamma^{2}_{s}+(\Delta_{p}+x+y)^{2}]$
whereas $\Delta_{p}=\nu - \omega_{L}$ while $\bar \Gamma_{s}=4\bar \Gamma_{0} + \bar \Gamma_{+} + \bar \Gamma_{-}$.
Figure (\ref{fig3}) shows the steady-state behavior of the linear susceptibility when the molecular sample is probed with a weak coherent field of frequency $\nu$. 
Both, positive or negative dispersions without absorption are clearly visible around $\Delta_{p}/\gamma=0$ which may lead to enhanced or reduced refractive 
indices applicable for optical imaging, lithography or negative refraction processes in dense media \cite{ref}.
In particular, the index of refraction close to vanishing absorption $n(\nu) \approx \sqrt{1+ \chi^{'}}$ takes  values $n >2$ for $\bar N =10^{17}{\rm cm^{-3}}$ 
and $\Delta/(2\Omega)=-0.43$. These dependences for the susceptibility $\chi(\nu)$ are easily understood in the double 
dressed-state picture (see Fig.~\ref{fig}c). Particularly, for $\Delta/(2\Omega)=0.43$ there are more molecules in the upper double dressed state 
$|\tilde 2\rangle$, i.e. $\langle \tilde R_{z}\rangle >0$ (see Fig.~\ref{fig2}b). As a consequence, at frequency $\omega_{L} + \omega + 2\bar G_{R}$ 
one has gain while at frequency $\omega_{L} - \omega - 2\bar G_{R}$ we have absorption, see the lateral dip/peak in Figure~(\ref{fig3}a). Similarly one can 
explain the whole structure shown in Figure~\ref{fig3}(a,b). An interesting issue here is the structure at $\omega_{L} \pm 2\bar G_{R}$ which may help 
to extract the value of permanent dipoles, i.e. $|d_{22}-d_{11}|$. If one inspects the absorption spectrum shown in Fig.~{\ref{fig3}} then the frequency 
separation between the first maximum and minimum around $\Delta_{p}=0$ is equal to $4\bar G_{R}$. $\bar G_{R}$ involves the difference of permanent 
dipoles $|d_{22}-d_{11}|$. This, in principle, allows to determine $|d_{22}-d_{11}|$ if all other involved parameters are known.
Finally, the elastic photon scattering spectrum consists of three lines at $\{ \omega_{L}, \omega_{L} \pm \omega \}$. 
The inelastic one may contain up to nine spectral lines in strict concordance with the double dressed-state formalism schematically shown in 
Fig.~(\ref{fig}c). Suppression of a spectral line at the frequency of the strongly driving laser also occurs \cite{suppr}. As well, the emission spectrum permits 
determination of the diagonal dipole matrix elements (see also Ref.~\cite{chk}).  The magnitudes of permanent dipoles are required for e.g. interpretation of 
biological images as well as for biological high-harmonic or ultrashort pulse generation processes via laser pumped media possessing permanent dipoles \cite{hhg} (see also \cite{wm}). 

\section{Summary}
Summarizing, we have investigated the steady-state quantum dynamics of laser pumped two-level molecular samples with broken inversion symmetry.
We have demonstrated population inversion in the bare states due to induced coherences which in turn depend on the magnitude of permanent dipoles.
Vacuum mediated collective effects among the two-level emitters considerably enhance the molecular inversion.
The values of the permanent dipoles can be inferred from the emission or absorption spectra. Furthermore, the investigated system exhibits large positive 
or negative dispersion without absorption facilitating applications including optical imaging, lithography and negative indices of refraction. The results apply 
especially to biomolecular, spin or asymmetrical quantum dot systems.
\acknowledgments
M.M. and C.H.K. acknowledge the financial support by the German Federal Ministry 
of Education and Research, grant No. 01DK13015, and Academy of 
Sciences of Moldova, grant No. 13.820.05.07/GF.  M.M. and M.M. are 
grateful for the nice hospitality of the Theory Division of the Max Planck Institute 
for Nuclear Physics from Heidelberg, Germany.



\begin{thebibliography}{55}
\bibitem{townes} A. L. Schawlow and C. H. Townes, Phys. Rev. {\bf 112}, 
1940 (1958).

\bibitem{prohorov} A. M. Prokhorov, JETP {\bf 34}, 1656 (1958).

\bibitem{basov} N. G. Basov, O. N. Krokhin, and Yu. M. Popov,
Sov. Phys. Usp. {\bf 3}, 702 (1961).

\bibitem{inv_at} C. M. Savage, Phys. Rev. Lett. {\bf 60}, 1828 (1988); 
M. H. Anderson, G. Vemuri, J. Cooper, P. Zoller and S. J. Smith, 
Phys. Rev. A {\bf 47}, 3202 (1993).
S. John and T. Quang, Phys. Rev. Lett. {\bf 78}, 1888 (1997); 
S. Hughes and H. J. Carmichael, {\it ibid.} {\bf 107}, 193601 (2011);
T. Quang and H. Freedhoff, Phys. Rev. A {\bf 47}, 2285 (1993).

\bibitem{inv_qd} T. M. Stace, A. C. Doherty and S. D. Barrett, Phys. Rev. Lett. {\bf 95}, 106801 (2005);
M. Gl\"{a}ssl, A. M. Barth and V. M. Axt, {\it ibid.} {\bf 110}, 147401 (2013); 
S. Das and M. A. Macovei, Phys. Rev. B {\bf 88}, 125306 (2013);
J. H. Quilter, A. J. Brash, F. Liu, M. Gl\"{a}ssl, A. M. Barth, V. M. Axt, A. J. Ramsay, M. S. Skolnick and A. M. Fox,
arXiv:1409.0913v3 [cond-mat.mes-hall].

\bibitem{lwi} O. A. Kocharovskaya and Ya. I. Khanin, JETP Lett. {\bf 48}, 630 (1988); 
S. E. Harris, Phys. Rev. Lett. {\bf 62}, 1033 (1989);  
M. O. Scully, S.-Y. Zhu and A. Gavrielides, {\it ibid.} {\bf 62}, 2813 (1989).

\bibitem{joerg} K. P. Heeg, H.-Ch. Wille, K. Schlage, T. Guryeva, D. Schumacher, I. Uschmann, K. S. Schulze, B. Marx, 
T. K\"{a}mpfer, G. G. Paulus, R. R\"{o}hlsberger and J. Evers, Phys. Rev. Lett. {\bf 111}, 073601 (2013);
K. P. Heeg, J. Haber, D. Schumacher, L. Bocklage, H.-Ch. Wille, K. S. Schulze, R. Loetzsch, I. Uschmann, G. G. Paulus, R. R\"{u}ffer, 
R. R\"{o}hlsberger, and J. Evers, {\it ibid.} {\bf 114}, 203601 (2015); K. P. Heeg, C. Ott, D. Schumacher, H.-C. Wille, 
R. R\"{o}hlsberger, T. Pfeifer, and J. Evers, {\it ibid.} {\bf 114}, 207401 (2015).


\bibitem{q_bio} N. Lambert, Y.-N. Chen, Y.-C. Cheng, C.-M. Li, G.-Y. Chen and F. Nori, Nature Phys. {\bf 9}, 10 (2013).

\bibitem{q_coh} G. S. Engel, Procedia Chemistry {\bf 3}, 222 (2011).

\bibitem{q_tr1} G. S. Engel, T. R. Calhoun, E. L. Read, T.-K. Ahn, T.  Mancal, Y.-C. Cheng, R. E. Blankenship and G. R. Fleming,
Nature (London) {\bf 446}, 782 (2007).

\bibitem{q_tr2} R. Hildner, D. Brinks, J. B. Nieder, R. J. Cogdell and N. F. van Hulst,
Science {\bf 340}, 1448 (2013).

\bibitem{lq_coh} E. Collini and G. D. Scholes, Science {\bf 323}, 369 (2009).

\bibitem{exp_rec} T. Ritz, P. Thalau, J.B. Phillips, R. Wiltschko and W. Wiltschko,
Nature (London) {\bf 429}, 177 (2004).

\bibitem{plenio} J. Cai and M. B. Plenio, Phys. Rev. Lett. {\bf 111}, 230503 (2013).

\bibitem{m_rec} E. M. Gauger, E. Rieper, J. J. L. Morton, S. C. Benjamin and V. Vedral, Phys. Rev. Lett. {\bf 106}, 040503 (2011); 
V. Vedral, Procedia Chemistry {\bf 3}, 172 (2011).

\bibitem{sms} H. P. Lu, L. Xun and  X. S. Xie, Science {\bf 282}, 1877 (1998).

\bibitem{ph_st} Y. Zheng and F. L. H. Brown, Phys. Rev. Lett. {\bf 90}, 238305 (2003).

\bibitem{nj} J. Eckel, J. H. Reina and M. Thorwart, New Jr. of Phys. {\bf 11}, 085001 (2009).

\bibitem{kov} V. A. Kovarskii, Phys. Usp. {\bf 42}, 797 (1999).

\bibitem{koch} M. O. Scully, Phys. Rev. Lett. {\bf 67}, 1855 (1991); 
Ch. O'Brien, P. M. Anisimov, Yu. Rostovtsev, and O. Kocharovskaya, Phys. Rev. A {\bf 84}, 063835 (2011).

\bibitem{setl} R. B. Setlow, E. C. Pollard, {\it Molecular Biophysics} (Oxford: Pergamon Press, 1962).

\bibitem{spin} A. S. M. Windsor, C. Wei, S. A. Holmstrom, J. P. D. Martin and N. B. Manson,
Phys. Rev. Lett. {\bf 80}, 3045 (1998).

\bibitem{q_dots} O. V. Kibis, G. Ya. Slepyan, S. A. Maksimenko and A. Hoffmann, Phys. Rev. Lett. {\bf 102}, 023601 (2009);
F. Oster, C. H. Keitel and M. Macovei, Phys. Rev. A {\bf 85}, 063814 (2012); E. Paspalakis, J. Boviatsis and S. Baskoutas, 
J. Appl. Phys. {\bf 114}, 153107 (2013).

\bibitem{gong} M. Terauchi, and T. Kobayashi, Chem. Phys. Lett. {\bf 137}, 319 (1987); 
A. Brown, and W. J. Meath, J. Chem.  Phys. {\bf 109}, 9351 (1998);
Ch. Brunel, B. Lounis, Ph. Tamarat, and M. Orrit, Phys. Rev. Lett. {\bf 81}, 2679 (1998);
V. Puller, B. Lounis, and F. Pistolesi, {\it ibid.} {\bf 110}, 125501 (2013); 
W. Yang, Sh. Gong, and Zh. Xu, Opt. Express {\bf 14}, 7216 (2006); 
F. Herrera, B. Peropadre, L. A. Pachon, S. K. Saikin, A. Aspuru-Guzik,
arXiv:1409.1930v1 [quant-ph].

\bibitem{gsa_book} G. S. Agarwal, {\it Quantum Statistical Theories of 
Spontaneous Emission and their Relation to other Approaches} 
(Springer, Berlin, 1974).

\bibitem{al_eb} L. Allen, J. H. Eberly, {\it Optical Resonance and 
Two-Level Atoms} (Dover, New York, 1975).

\bibitem{puri} R. R. Puri, {\it Mathematical Methods of Quantum Optics}
(Springer, Berlin 2001).

\bibitem{martin} M. Kiffner, M. Macovei, J. Evers, and C. H. Keitel, Progress in
Optics {\bf 55}, 85 (2010).

\bibitem{ref} N. A. Proite, B. E. Unks, J. T. Green, and D. D. Yavuz, Phys. Rev. Lett. {\bf 101}, 147401 (2008);
Z. J. Simmons, N. A. Proite, J. Miles, D. E. Sikes, and D. D. Yavuz, Phys. Rev. A {\bf 85}, 053810 (2012).

\bibitem{suppr} Y. He {\it et al.}, Phys. Rev. Lett. {\bf 114}, 097402 (2015).

\bibitem{chk} O. Postavaru, Z. Harman, and C. H. Keitel, Phys. Rev. Lett. {\bf 106}, 033001 (2011).

\bibitem{hhg} V. A. Kovarsky, B. S. Philipp, and E. V. Kovarsky, Phys. Lett. A {\bf 226}, 321 (1997); 
W. Yang, Sh. Gong, R. Li, and Zh. Xu, {\it ibid.} {\bf 362}, 37 (2007).

\bibitem{wm} J. Deiglmayr, A. Grochola, M. Repp, O. Dulieu, R. Wester, and M. Weidem\"{u}ller, 
Phys. Rev. A {\bf 82}, 032503 (2010).
\end{thebibliography}
\end{document}